\documentclass[12pt]{article}
\usepackage{graphicx}

\begin{document}

\begin{center}
\Large {\bf  Entanglement
R\'{e}nyi Entropies in Conformal Field Theories and Holography}
\end{center}

\bigskip
\bigskip

\begin{center}
D.V. Fursaev
\end{center}

\bigskip
\bigskip

\begin{center}
{\it Dubna International University \\
     Universitetskaya str. 19\\
     141 980, Dubna, Moscow Region, Russia\\

  and\\

  the Bogoliubov Laboratory of Theoretical Physics\\
  Joint Institute for Nuclear Research\\
  Dubna, Russia\\}
 \medskip
\end{center}

\bigskip
\bigskip

\begin{abstract}
An entanglement R\'{e}nyi entropy for a spatial partition of a system is studied
in conformal theories which admit a dual description in terms of an anti-de Sitter
gravity. The divergent part of the R\'{e}nyi entropy is computed in 4D conformal ${\cal N}=4$
super Yang-Mills theory at a weak coupling. This result is used to suggest
a holographic formula  which reproduces the R\'{e}nyi entropy at least in the leading approximation.
The holographic R\'{e}nyi entropy is an invariant functional set on a codimension 2 minimal hypersurface
in the bulk geometry.  The bulk space does not depend on order $n$ of the R\'{e}nyi entropy.
The holographic R\'{e}nyi entropy is a sum of local and non-local  functionals
multiplied by polynomials of $1/n$.
\end{abstract}

\newpage

\section{Introduction}

A holographic description of many-body systems, including quantum field  and condensed matter
theories, in terms of gravity theories one dimension higher is an active research area.
One of the aims here is to get new insights
in situations where traditional
methods meet difficulties, in regimes of strong couplings, for a example.

Quantum entanglement is one of those notions which carries an information about strength of
correlations
in a system. If a quantum system specified by a density matrix  $\hat{\rho}$ is divided
spatially onto parts, $A$ and $B$,
one can define a reduced density matrix, say, for the region $A$,
\begin{equation}\label{2.6}
\hat{{\rho}}_A=\mbox{Tr}_B \hat{\rho}~~,
\end{equation}
by taking trace over the states located in the region $B$.
To quantify the degree of entanglement one introduces the entanglement entropy
\begin{equation}\label{2.1}
S_A=-\mbox{Tr}_A \hat{{\rho}}_A\ln \hat{{\rho}}_A~~,
\end{equation}
and the entanglement R\'{e}nyi entropy of order $n$
\begin{equation}\label{2.2}
S^{(n)}_A={\ln \mbox{Tr}_A \hat{{\rho}}^{~n}_A \over 1-n}~~,
\end{equation}
where $n=2,3,...$. Formally $S_A=\lim_{n\to 1}S^{(n)}_A$.

In seminal papers \cite{Ryu:2006bv,Ryu:2006ef} Ryu and Takayanagi  suggested a "holographic formula"
for calculation of entanglement entropy in conformal field theories (CFT) which admit a dual description
in terms of anti-de Sitter (AdS) gravity. For a $d$ dimensional CFT spatially divided
by a surface $\cal B$
the corresponding  entropy  of entanglement between the two parts
is given by the Bekenstein-Hawking-like formula
\begin{equation}\label{1.1}
S(\tilde{\cal B})= {\mbox{vol}(\tilde{\cal B}) \over 4G_N^{(d+1)}}~~~.
\end{equation}
Here $G_N^{(d+1)}$ is a gravitational constant in a dual
gravity theory and $\mbox{vol}(\tilde{\cal B})$ is the volume of a  certain
codimension 2 hypersurface $\tilde{\cal B}$ lying in the bulk. The definition of
$\tilde{\cal B}$ is a classical Plateau problem:
$\tilde{\cal B}$ has the least volume
among the codimension 2 hypersurfaces in the bulk whose asymptotic infinity belongs
to a conformal class of $\cal B$. There are extra, topological, requirements \cite{Fursaev:2006ih}
for  the choice
of $\tilde{\cal B}$ to distinguish between cases when the entropy corresponds to the reduced matrix
$\hat{{\rho}}_A$ or $\hat{{\rho}}_B$.

Formula (\ref{1.1}) passes non-trivial tests. It reproduces known
explicit  expressions obtained by direct computations  in 2D and 4D CFT's.
Among recent interesting
applications of (\ref{1.1}) are works on critical phenomena \cite{Klebanov:2007ws}, higher
dimensional extensions of the $c$-theorems \cite{Hung:2011xb}-\cite{Myers:2010tj}, boundary effects
in entanglement entropy \cite{Takayanagi:2011zk} and many others, see
\cite{Nishioka:2009un} for a general review, and \cite{Solodukhin:2011gn} for a possible
role of entanglement in the origin of the entropy of black holes.

It is not much known about entanglement R\'{e}nyi entropy (ERE) (\ref{2.2}) in field theory
models and about its holographic representation.
An extensive analysis, mainly in 2D CFT's, for two disjoint intervals was done in
\cite{Headrick:2010zt}. In \cite{Casini:2010kt} the logarithmic part of ERE was obtained for
a massless scalar field in Minkowsky space-time and spherical entangling
surface $\cal B$. The idea of \cite{Casini:2010kt} is that in the given example the reduced density
matrix $\hat{{\rho}}_A$, see (\ref{2.6}), can be converted in a thermal density matrix.
This method was applied in \cite{Klebanov:2011uf} to get ERE for free scalar and spinor fields
in 3 dimensions. The same idea was used in \cite{Hung:2011nu} to calculate ERE in various
holographic models.

The definition of a holographic ERE may require a smooth modification of the bulk geometry with dependence
of the bulk metric on order $n$. This possibility was explored in \cite{Hung:2011nu}
by identifying ERE in an effective thermal state in the boundary CFT with an entropy of a
black hole in the bulk.
In our work we study  another option by assuming that the bulk
geometry in the definition of the holographic ERE
does not depend on $n$. We calculate leading terms of
the entanglement R\'{e}nyi entropy (\ref{2.2}) and use this information to
suggest a corresponding generalization of the Ryu-Takayanagi formula.

Our computations
done in the limit
of the weak coupling are summarized by the formula
\begin{equation}\label{1.2}
S^{(n)}({\cal B})\simeq \sum_{p=2}^{d-1}\Lambda^{d-p}{s^{(n)}_{~p} \over d-p}+s^{(n)}_{~d}\ln(\Lambda\mu)+...~~,
\end{equation}
where  $d$ is the dimensionality of the space-time (a boundary theory),
$\Lambda$ is an ultraviolet cutoff,
$\mu$ is a typical scale of the theory. The canonical mass dimensions of  $\Lambda$ and $\mu$ are
$+1$ and $-1$, respectively. We assume that the
space and, consequently, ${\cal B}$ have no boundaries.
One can show then that $s^{(n)}_{~p}=0$ for odd $p$.

Our result for 4D CFT (${\cal N}=4$ $SU(N)$ supersymmetric Yang-Mills theory) is that $s^{(n)}_{~2}$
is proportional to $\gamma_n A({\cal B})$, where $A({\cal B})$ is the area of ${\cal B}$ and $\gamma_n=n^{-1}$.
The coefficient $s^{(n)}_{~4}$ is related to the conformal anomaly. It is
a scale invariant functional of the following structure:
\begin{equation}\label{1.3}
s^{(n)}_{~4}=d(N)\left( a(\gamma_n)F_a+c(\gamma_n)F_c+b(\gamma_n)F_b\right)~~,
\end{equation}
where $d(N)=N^2-1$, $F_a$ is related to the Euler characteristic of ${\cal B}$, $F_c$
is an integral of a projection of the Weyl
tensor on ${\cal B}$, and $F_b$ is constructed solely of the extrinsic curvatures of ${\cal B}$.
Coefficients $c(\gamma_n)$ and $a(\gamma_n)$ are 3d order polynomials of $\gamma_n$ which we compute
by methods of the spectral geometry. Our method does not allow one to fix $b(\gamma_n)$.

It is the structure of  Eqs. (\ref{1.2}), (\ref{1.3}) which motivates our
suggestion of a holographic ERE. The main result here is formula (\ref{7.21}) which reproduces
(\ref{1.2}), (\ref{1.3}). The holographic ERE coincides with Ryu-Takayanagi formula (\ref{1.1})
at $n=1$ but it has a more complicated structure, in general.
The holographic R\'{e}nyi entropy (\ref{7.21}) in $d=4$ is
an invariant functional set on a codimension 2 minimal hypersurface $\tilde{\cal B}$ in the bulk
which does not depend on $n$.
Similar to (\ref{1.3}) the entropy functional is linear combination
of polynomials related to $a(\gamma_n)$, $b(\gamma_n)$, $c(\gamma_n)$ and 4 different invariants.
One of the invariants is $\mbox{vol}(\tilde{\cal B})$,
two other are analogous to $F_c$ and $F_b$ and expressed with the help of curvatures in the bulk.
The functional corresponding to $F_a$ is non-local logarithmic correction $\ln\mbox{vol}(\tilde{\cal B})$.

The work is organized as follows. Computations of the UV part of ERE in
2D and 4D CFT's along with derivation of (\ref{1.2}), (\ref{1.3}) are presented in Sec. \ref{RCFT}.
The holographic entanglement R\'{e}nyi entropy is suggested in Sec. \ref{HRE}.
Sec. \ref{spec} contains a discussion of the results and concluding remarks.
In particular, we  speculate here on how the holographic
formulas for ERE and for the entanglement entropy may appear in quantum
gravity.
The proposal of Sec. \ref{HRE} is based on asymptotic behavior of different
curvature invariants near the AdS boundary. Proofs of corresponding mathematical statements, some
of which are new, are collected in Appendices. We show in Appendix B that the tilt angle
of a minimal hypersurface $\tilde{\cal B}$ near a boundary of an asymptotically AdS space is determined
by the extrinsic curvature of $\cal B$.
This enables one to derive asymptotic embedding equations of $\tilde{\cal B}$
from pure geometrical considerations, see Appendix C.

\section{R\'{e}nyi entropy in conformal theories}\label{RCFT}
\setcounter{equation}0

\subsection{Basic definitions}

We consider a conformal field theory (CFT) set on a $(d-1)$-dimensional constant time hypersurface $\Sigma$.
The spacetime is assumed to be static. The CFT density matrix $\hat{\rho}$ is chosen to be thermal,
$\hat{\rho}=e^{-\hat{H}/T}/Z(T)$, where $T$ is the temperature, $\hat{H}$ is a Hamiltonian,
and $Z(T)=\mbox{Tr}~\exp(-\hat{H}/T)$ is a partition function of the CFT.

It is convenient to introduce an 'entanglement partition function' (EPF)
associated to division of $\Sigma$ onto the parts $A$ and $B$ by a
surface $\cal B$,
\begin{equation}\label{2.5}
Z(n,T)=\mbox{Tr}_A\left(\mbox{Tr}_B~e^{-\hat{H}/T}\right)^n~~,
\end{equation}
where $n$ are natural numbers.
At $n=1$ the two partition functions coincide, $Z(1,T)=Z(T)$.
As follows from (\ref{2.6}) and (\ref{2.5}), the entanglement entropies can be expressed as
\begin{equation}\label{2.3}
S(T)=-\lim_{n \rightarrow 1}~ \left(n {\partial \over
\partial n}-1\right) \ln Z(n,T)~~~,
\end{equation}
\begin{equation}\label{2.4}
S^{(n)}(T)={1\over 1-n}\left(\ln Z(n,T)-n \ln Z(T)\right)~~.
\end{equation}
In (\ref{2.3}) one takes the limit $n \to 1$ by going from discrete to a continuous $n$.
Arguments in support of this procedure can be found in  \cite{Nesterov:2010yi}.
We imply but omit in (\ref{2.5}), (\ref{2.3}), (\ref{2.4}) the index $A$ (compare with
(\ref{2.1}), (\ref{2.2})). The ground state EPF, $Z(n)$, can be obtained
from $Z(n,T)$ in the limit $T\to 0$.

In a quantum field theory the partition function $Z(T)$
is represented as a functional integral over field configurations
which live on a Euclidean static $d$ dimensional manifold
${\cal M}$ with the constant time sections $\Sigma$. The
orbits of the Killing vector field generating translations in Euclidean time
are the circles $S^1$ with the length equal $1/T$.

Analogously, $Z(n,T)$ can be written in terms of a path integral
where field configurations
are set on a 'replicated' manifold ${\cal M}_n$ which is glued from $n$ copies (replicas) of $\cal M$
along some cuts which meet on $\cal B$. An explicit construction of ${\cal M}_n$
is described in \cite{Fursaev:2006ng}. For our purpose it is enough to know that ${\cal M}_n$
are locally identical to $\cal M$ but have nontrivial topologies: ${\cal M}_n$ have conical singularities
on $\cal B$ with the length of a small unit circle around each point on $\cal B$ equal $2\pi n$.

The partition function in a free CFT, therefore, is
\begin{equation}\label{6.1}
-\ln Z(n)=\frac 12 \sum_i\eta_i\ln \det (\mu^2 \Delta^{(i)})~~,
\end{equation}
where $\Delta^{(i)}$ are Laplace operators for different fields which enter the model, $\eta_i=+1$
for Bosons and $\eta_i=-1$ for Fermions, $\mu$ is a scale parameter. The base manifold for the
Laplace operators
is ${\cal M}_n$. Determinant of an operator $L$ can be defined, for example, by the Ray-Singer formula:
$\ln \det L=-\zeta'(0;L)$ in terms of a first derivative of the zeta-function $\zeta(s;L)$ of $L$.
For our purposes we use only
scalar, spinor and vector Laplacians which are, respectively, $\Delta^{(0)}=-\nabla^2+\xi R$,
$\Delta^{(1/2)}=(i\gamma^\mu\nabla_\mu)^2$, $(\Delta^{(1)})^\nu_\mu=-\nabla^2\delta^\nu_\mu+R^\nu_\mu$.
Here $R$, $R^\nu_\mu$ are the scalar curvature and the Ricci tensor.
Quantization of vector fields is done in the Lorentz gauge and  produces a couple of ghost fields.
The Laplacians $\Delta^{(0)}$ for ghosts have minimal coupling, $\xi=0$.

When the Laplace operators do not have zero eigenvalue modes the ultraviolet part of (\ref{6.1}) is
\begin{equation}\label{6.2}
-\ln Z(n)\simeq \sum_{p=0}^{d-1}\Lambda^{d-p}~{A_{p}(n) \over p-d}-A_{d}(n)\ln(\Lambda\mu)~~,
\end{equation}
\begin{equation}\label{6.3}
A_p(n)\equiv \sum_i\eta_iA_p(\Delta^{(i)})~~,
\end{equation}
where $A_p(\Delta^{(i)})$ are the heat coefficients that appear in short $t$
expansions for the corresponding heat kernel operators on ${\cal M}_n$,
\begin{equation}\label{6.4}
\mbox{Tr}~e^{-t\Delta^{(i)}}\simeq \sum_p A_p(\Delta^{(i)})~t^{(p-d)/2}~~.
\end{equation}
If there are no boundaries
$A_p(\Delta^{(i)})=0$ for odd $p$. The heat coefficients have been computed
earlier by different authors.
We give corresponding references and describe the coefficients more carefully 
in sec. \ref{4DCFT},
see Eqs. (\ref{6.17}), (\ref{6.20}).
Equation (\ref{1.2}) follows from (\ref{2.4}), (\ref{6.2})
if one puts
\begin{equation}\label{6.5}
s^{(n)}_{~p} ={nA_p(1)-A_p(n) \over n-1}~~.
\end{equation}
The Ray-Singer definition takes into account only non-zero eigenvalues of an operator.
Therefore, when a Laplace operator $\Delta^{(i)}$ has
a certain number $N_{\mbox{\tiny{zm}}}^{(i)}$ of zero modes
one should use in (\ref{6.2}) a combination $A_d(\Delta^{(i)})-N_{\mbox{\tiny{zm}}}^{(i)}$.
This results in the following modification of (\ref{6.5}) for the partial entropy:
\begin{equation}\label{6.5a}
s^{(n)}_{~p=d} ={nA_d(1)-A_d(n) \over n-1}+s^{(n)}_{\mbox{\tiny{zm}}}~~,
\end{equation}
\begin{equation}\label{6.5b}
s^{(n)}_{\mbox{\tiny{zm}}} =-{nN_{\mbox{\tiny{zm}}}(1)-N_{\mbox{\tiny{zm}}}(n) \over n-1}~~,
\end{equation}
\begin{equation}\label{6.5c}
N_{\mbox{\tiny{zm}}}(n)=\sum_i\eta_iN_{\mbox{\tiny{zm}}}^{(i)}~~.
\end{equation}
The simplest example is a Laplace operator on a compact 2D manifold $\cal M$. Such an operator on  $\cal M$
and on the corresponding replicated spaces ${\cal M}_n$, which are compact as well,
has a single normalizable zero mode. Hence $s^{(n)}_{~p=2}=-1$.
In general, the number of zero modes (and so $s^{(n)}_{~p=d}$)
may depend on the order $n$.
In what follows we ignore effects of zero modes.

\subsection{2D CFT}

A simplest 2D CFT consists of some number of free spinor and minimally coupled scalar fields.
The R\'{e}nyi entropy is given by (\ref{1.2}) for $d=2$. There is only a logarithmic term.
From (\ref{6.5}) one gets the known result \cite{Calabrese:2004eu}
\begin{equation}\label{6.6a}
s^{(n)}_{~2}={c \over 12}(1+\gamma_n){\cal A}({\cal B})~~,
\end{equation}
where $c$ is the total number of fields (the sum of central charges),
${\cal A}({\cal B})$ is the number of points of ${\cal B}$
(for example, $A({\cal B})=2$ if ${\cal B}$ is an interval).

\subsection{4D CFT}\label{4DCFT}

${\cal N}=4$ $SU(N)$ supersymmetric Yang-Mills theory in $d=4$  consists of 6 multiplets
of conformally coupled scalar fields (with $\xi=1/6$), 4 multiplets of Weyl spinors, and 1 multiplet
of gluon fields.
Each multiplet is in adjoint representation of the $SU(N)$ group.
Computations in the case of the zero coupling with the help of (\ref{6.5}) yield
\begin{equation}\label{6.6}
s^{(n)}_{~2}={d(N) \over 4\pi}\gamma_n {\cal A}({\cal B})~~,
\end{equation}
\begin{equation}\label{6.8}
s^{(n)}_{~4}=d(N)\left(a(\gamma_n)F_a+c(\gamma_n)F_c+b(\gamma_n)F_b\right)~~,
\end{equation}
where $d(N)=N^2-1$,
\begin{equation}\label{6.9}
F_a=-{1 \over 2\pi}\int_{{\cal B}}\sqrt{\sigma}d^2x~R({\cal B})=-2\chi~~,
\end{equation}
and $\chi$ is the Euler characteristic of ${\cal B}$ (since ${\cal B}$ is closed and has topology
of $S^2$, hence $\chi=2$),
$a(\gamma_n)$ and $c(\gamma_n)$ are the following polynomials:
\begin{equation}\label{6.11}
a(\gamma_n)={1 \over 96}\left(\gamma_n^3+\gamma_n^2+7\gamma_n+15\right)~~,
\end{equation}
\begin{equation}\label{6.10}
c(\gamma_n)={1 \over 32}\left(\gamma_n^3+\gamma_n^2+3\gamma_n+3\right)~~.
\end{equation}
The functional $F_c$ is determined in terms of a projection  $C_{ijij}$
of the Weyl tensor  $C_{\mu\nu\lambda\rho}$ at $\cal B$,
\begin{equation}\label{6.12}
F_c={1 \over 2\pi}\int_{{\cal B}}\sqrt{\sigma}d^2x~C_{ijij}~~,
\end{equation}
\begin{equation}\label{6.13}
C_{ijij}=C_{\mu\nu\lambda\rho}n_i^\mu n_j^\nu n_i^\lambda n_i^\rho~~.
\end{equation}
$n_i$, $i=1,2$, are two unit mutually orthogonal outward pointing normal vectors to ${\cal B}$.
The Weyl tensor is
$$
C_{\mu\nu\lambda\rho}=R_{\mu\nu\lambda\rho}+{1 \over d-2}\left(g_{\mu\rho}R_{\nu\lambda}
+g_{\nu\lambda}R_{\mu\rho}-g_{\mu\lambda}R_{\nu\rho}-g_{\nu\rho}R_{\mu\lambda}\right)
$$
\begin{equation}\label{6.14}
+{R \over (d-1)(d-2)}\left(g_{\mu\lambda}g_{\nu\rho}-g_{\mu\rho}g_{\nu\lambda}\right)~~,
\end{equation}
where $R$, $R_{\mu\nu}$,
$R_{\mu\nu\lambda\rho}$ are, respectively, the scalar curvature, the Ricci tensor and the Riemann tensor
of $\cal M$. Finally,
\begin{equation}\label{6.15}
F_b={1 \over 2\pi}\int_{{\cal B}}\sqrt{\sigma}d^2x~\left({1 \over (d-2)} k_i^2-\mbox{Tr}(k_i^2)\right)~~,
\end{equation}
where $(k_i)_{\mu\nu}=h_{\mu}^\lambda h_{\nu}^\rho (n_i)_{\mu;\nu}$ are extrinsic curvatures of $\cal B$,
$k_i=g^{\mu\nu}(k_i)_{\mu\nu}$, $\mbox{Tr}(k_i^2)=(k_i)_{\mu\nu}(k_i)^{\mu\nu}$.
Note that $d=4$ in (\ref{6.14}), (\ref{6.15}) for theories in 4 dimensions.

The functionals $F_c$, $F_b$ (and certainly $F_a$) are invariant
under conformal transformations of the metric
\begin{equation}\label{6.16}
\bar{g}_{\mu\nu}(x)=e^{-2\omega(x)}g_{\mu\nu}(x)~~,~~\bar{n}_i^\mu=e^{\omega}n_i^\mu~~.
\end{equation}
Conformal transformations are discussed in Appendix A. The conformal invariance is a consequence of the
properties of the heat coefficients.

Coefficient functions $c(\gamma_n)$, $a(\gamma_n)$, $b(\gamma_n)$ are related to the
contribution of conical singularities
to the heat coefficients of Laplace type operators on singular base manifolds. The coefficients
have the following structure:
\begin{equation}\label{6.17}
A_p(\Delta^{(i)})=nA_p(\Delta^{(i)})_{n=1}+\bar{A}_p(\Delta^{(i)})~~.
\end{equation}
The term related to the presence of the conical singularities, $\bar{A}_p(\Delta^{(i)})$, is proportional
to $(1-n)$ and can be written as
\begin{equation}\label{6.18}
\bar{A}_p(\Delta^{(i)})=(1-n)\eta_is^{(n)}_{~p,i}~~.
\end{equation}
It follows from (\ref{6.18}) that $s^{(n)}_{~p,i}$ is a contribution to the
$p$-th term of the R\'{e}nyi entropy from a particular field
\begin{equation}\label{6.19}
s^{(n)}_{~p} =\sum_i s^{(n)}_{~p,i}~~,
\end{equation}
see (\ref{6.5}).
If the singular part of the heat coefficient is represented as
\begin{equation}\label{6.20}
\bar{A}_4(\Delta)=\bar{a}(\gamma_n)F_a+\bar{c}(\gamma_n)F_c+\bar{b}(\gamma_n)F_b
\end{equation}
calculations in four dimensions yield for $\bar{a}(\gamma_n)$ and $\bar{c}(\gamma_n)$
the values which are summarized in Table \ref{t1}.
For a gauge boson the given result takes into account a contribution of ghosts.
For the sake of clarity we also give
the relation between coefficients defined in (\ref{6.11}), (\ref{6.10}) and
coefficients from Table \ref{t1}
\begin{equation}
\label{rel1}
a(\gamma_n)={1 \over 1-n}(6\bar{a}_{0}(\gamma_n)
-4\bar{a}_{1/2}(\gamma_n)+\bar{a}_{1}(\gamma_n))~~,
\end{equation}
\begin{equation}
\label{rel2}
c(\gamma_n)={1 \over 1-n}(6\bar{c}_{0}(\gamma_n)
-4\bar{c}_{1/2}(\gamma_n)+\bar{c}_{1}(\gamma_n))~~,
\end{equation}
where indexes $0,1/2$ and $1$ correspond to scalar, Weyl spinor and vector fields, respectively.

Computations of coefficient $A_2$
for spin 1/2 and 1 can be found in
\cite{Kabat:1995eq} and \cite{Fursaev:1996uz} (along with references to spin 0 results).
Computations of $A_4$ can be found in different works: for spin 0 in \cite{Fursaev:1994},
\cite{Dowker:1994bj}, \cite{Dowker:1994nt}, for spin 1/2 in \cite{Fursaev:1997th},
and for spin 1 in \cite{DeNardo:1996kp}. Let us emphasize that all computations imply
that conical singularities
are located on a surface with vanishing extrinsic curvatures. Some information on the effect
of the curvatures can be obtained from requirement that $A_4$ in conformal theories in $d=4$ is scale
invariant, see, e.g. \cite{Fursaev:2011zz} for discussion of this property. As was pointed out in
\cite{Dowker:1994bj}, \cite{Dowker:1994nt} $A_4$ can be fixed up to adding some conformally invariant
functional of extrinsic curvatures. This functional, $F_b$, is introduced in  (\ref{6.15}). It is a
quadratic combination of the curvatures
because $A_4$ has zero canonical mass dimension in $d=4$.

The coefficient $\bar{b}(\gamma_n)$ has not been derived so far by a direct computation.
As we show later by using holographic
arguments of \cite{Solodukhin:2008dh}, $\bar{b}(\gamma_n=1)=1/4$. This means that function
$\bar{b}(\gamma_n)$ may be non-trivial. It should be mentioned that, if $F_b=0$,
formula (\ref{6.8}) can be used with unknown $\bar{b}(\gamma_n)$ even in cases of non-vanishing
extrinsic curvatures. An example is a spherical entangling
surface in a theory in Minkowsky space-time. One can easily find corresponding ERE for this model in $d=4$
by using (\ref{6.18}), (\ref{6.20}) and results of Table \ref{t1}.
Here $F_a=-4$, $F_c=F_b=0$ and one finds, in particular, that $s^{(n)}_4=-(\gamma_n+1)(\gamma_n^2+1)/360$ for scalars, 
$s^{(n)}_4=-(\gamma_n+1)(7\gamma_n^2+37)/1440$ for Weyl spinors.
Computations of the logarithmic ERE $s^{(n)}_4$
in this model have been done in \cite{Casini:2010kt},\cite{Klebanov:2011uf} by transforming the reduced
density matrix to a thermal form. The results of \cite{Casini:2010kt},\cite{Klebanov:2011uf}
completely agree with the result above. 

As for ERE for the spherical entangling surface 
in the weakly coupled supersymmetric Yang-Mills theory the same computation
yields
\begin{equation}
\label{rel3}
S^{(n)}(R)\simeq d(N)\left[ {\Lambda^2 \over 8\pi n}{\cal A}-
{1 \over 48n^3}(15n^3+7n^2+n+1)\ln(\Lambda^2{\cal A})\right]~~,
\end{equation}
where  ${\cal A}=4\pi R^2$ is the area of the surface and $R$ is its radius.
To write (\ref{rel3}) we used (\ref{1.2}), where the infrared cutoff parameter $\mu$ was replaced with
the radius $R$. This result disagrees with a holographic computation of the same entropy in \cite{Hung:2011nu}
where ERE was identified with the entropy of a black hole in the AdS gravity. 
The difference is in the dependence on $n$ (in \cite{Hung:2011nu} this dependence is
not the ratio of polynomials). We return to discussion of this point in sec. \ref{spec}.

\begin{table}
\renewcommand{\baselinestretch}{2}
\medskip
\caption{Coefficient functions of singular parts of the heat coefficients}
\bigskip
\begin{centerline}
{\small
\begin{tabular}{|c|c|c|c|c|}
\hline
$\mbox{field}$  & $\bar{c}(\gamma)$ & $\bar{a}(\gamma)$  & $c$ & $a$  \\
%         &             &              &   & &        \\
\hline
$\mbox{real scalar}$ &  ${\gamma^4-1 \over 480 \gamma}$
 &  ${\gamma^4-1 \over 1440 \gamma}$   & ${1 \over 120}$ & ${1 \over 360}$ \\
\hline
$\mbox{Weyl spinor}$ &  $-{7\gamma^4+10\gamma^2-17 \over 1920 \gamma}$
 &  $-{7\gamma^4+30\gamma^2-37 \over 5760 \gamma}$ & ${1 \over 40}$ & ${11 \over 720}$ \\
\hline
$\mbox{gauge Boson}$ &  ${\gamma^4+10\gamma^2-11 \over 240 \gamma}$
 &  ${\gamma^4+30\gamma^2+60\gamma-91 \over 720 \gamma}$ & ${1 \over 10}$ & ${31 \over 180}$ \\
\hline
\end{tabular}}
\bigskip
\renewcommand{\baselinestretch}{1}
\end{centerline}
\label{t1}
\end{table}

There is a relation of the functions $\bar{a}(\gamma_n)$, $\bar{c}(\gamma_n)$ to the conformal anomaly.
The trace of the renormalized
stress-energy tensor of the each field has the form
\begin{equation}\label{6.23}
\langle T^\mu_\mu \rangle=-a~E- c~ I~~,
\end{equation}
\begin{equation}\label{6.24}
E={1 \over 16\pi^2}\left(R_{\mu\nu\lambda\rho}R^{\mu\nu\lambda\rho}-4R_{\mu\nu}R^{\mu\nu}+R^2\right)~~,
\end{equation}
\begin{equation}\label{6.25}
I=-{1 \over 16\pi^2}C_{\mu\nu\lambda\rho}C^{\mu\nu\lambda\rho}~~.
\end{equation}
Constants $a$ and $c$ are given in Table \ref{t1} and one can check that
\begin{equation}\label{6.26}
c=\eta \partial_\gamma\bar{c}(\gamma=1)~~,~~a=\eta \partial_\gamma\bar{a}(\gamma=1)~~.
\end{equation}
(As earlier, $\eta=+1$ for Bosons and $\eta=-1$ for Fermions.)
On a regular manifold $A_4$ is composed of
integrals of $E$ and $I$. Each of these integrals can be
defined also on a singular manifold, if $|\gamma-1|$ is small and  terms $O((\gamma-1)^2)$
are neglected, see \cite{Fursaev:1995ef}.
Equation (\ref{6.26}) follows from a property established first in \cite{Fursaev:1994ea} for scalar Laplacians
with minimal coupling: up to terms proportional to $(\gamma-1)$
the coefficient $A_4$ in the heat trace asymptotic
of a Laplace operator on a manifold with conical singularities coincides with
the corresponding heat coefficient on a regular manifold obtained by 'smoothing' conical singularities.

Relation of scaling properties of the entanglement entropy to the trace anomaly is discussed in
\cite{Schwimmer:2008yh},\cite{Myers:2010tj}.

\begin{figure}[h]
\begin{center}
\includegraphics[height=3cm,width=6.8cm]{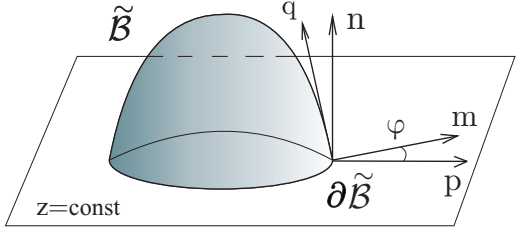}
%\v_space{0.4cm}
\caption{\small{The figure shows the holographic surface
$\tilde{\cal B}$ with normal vectors in a constant time section.}}
\label{F1}
\end{center}
\end{figure}

\section{Toward a holographic description of the R\'{e}nyi entropy}\label{HRE}
\setcounter{equation}0

\subsection{AdS gravity}

The AdS/CFT conjecture \cite{Ma}--\cite{GKP} states
that a supergravity theory in
the anti-de Sitter (AdS) space is dual to a conformal field theory on the boundary
of that region.
Thus, we consider the $d+1$ dimensional gravity theory
\begin{equation}\label{7.1}
I[g]=-{1 \over 16\pi G_N^{(d+1)}}\int_{\tilde{\cal M}} d^{d+1}x\sqrt{g}
\left(\tilde{R}+{d(d-1) \over l^2}\right)~~~
\end{equation}
with the negative cosmological constant
$-d(d-1)/(2l^2)$. In what follows we put $l=1$, for simplicity.

Let $\tilde{\cal M}$ be a manifold with a metric $g_{KL}$
which is a solution to the Einstein
equations in theory (\ref{7.1}). Since $\tilde{\cal M}$ is asymptotically AdS $I[g]$
is not well defined.
To avoid the volume divergencies of $I[g]$ one makes a cut of $\tilde{\cal M}$
at some $d$-dimensional hypersurface $\partial \tilde{\cal M}$. (We
imply in (\ref{7.1}) traditional boundary terms on $\partial \tilde{\cal M}$ but do not
write them explicitly.) The cut of $\tilde{\cal M}$ is determined in suitably chosen
coordinates by a fixed 'radius' $\varrho$.

Let $\cal M$ be a $d$ dimensional manifold
where a boundary CFT is defined on. The holographic relations
require that metric induced on the boundary $\partial \tilde{\cal M}$
belongs to the conformal class of ${\cal M}$ in the limit of the infinite radius $\varrho$.
The volume (infrared) divergences on the gravity side are identified
with ultraviolet divergences in the CFT, that is
$\varrho$ turns out to be related to a UV cutoff in the theory.

We denote scalar curvature, the Ricci tensor and the Riemann tensor of $\tilde{\cal M}$
by $\tilde{R}$, $\tilde{R}_{LP}$, $\tilde{R}_{LPKN}$, respectively.

\subsection{Geometrical structures at AdS asymptotic}

It is convenient to choose coordinates near $\partial \tilde{\cal M}$
in which the metric on $\tilde{\cal M}$ takes the form
\begin{equation}\label{7.1a}
ds^2=z^{-2}\left(dz^2 + g_{\mu\nu}(z,x)dx^\mu dx^\nu\right)~~.
\end{equation}

By definition, the embedding equation of $\partial \tilde{\cal M}$ is $z=const$. The relation with
the radius is $\varrho=1/z$.
If (\ref{7.1a}) is a solution to the bulk gravity equations the behavior of $g_{\mu\nu}(z,x)$ at small $z$
is known from the Fefferman-Graham asymptotic
\begin{equation}\label{7.2}
g_{\mu\nu}(z,x)=g_{\mu\nu}(x)+z^2g_{\mu\nu}^{(1)}(x)+...~~,
\end{equation}
see e.g. \cite{Henningson:1998gx}, where
\begin{equation}\label{7.3}
g_{\mu\nu}^{(1)}=-{1 \over d-2}\left(R_{\mu\nu}-{g_{\mu\nu} R \over 2(d-1)}\right)~~,
\end{equation}
$g_{\mu\nu}(x)$ is a metric of $\cal M$, $R_{\mu\nu}$ and  $R$ are curvatures of $\cal M$.

We assume that
manifold $\cal M$ is static, so does the solution $\tilde {\cal M}$. The constant time section of
$\tilde {\cal M}$ and its intersection with $\partial \tilde {\cal M}$ are denoted as
$\tilde{\Sigma}$ and $\partial \tilde{\Sigma}$, respectively. Constant time section $\Sigma$
of $\cal M$ and $\partial \tilde{\Sigma}$ belong to the same conformal class.

We consider a minimal codimension 2 hypersurface $\tilde{\cal B}$ lying in a constant
time section of $\tilde{\cal M}$. (The fact that the space is static implies
that $\tilde{\cal B}$ is also minimal in $\tilde {\Sigma}$.) $\tilde{\cal B}$ ends on
$\partial \tilde{\Sigma}$.
It is required that the boundary $\partial \tilde {\cal B}$ of $\tilde{\cal B}$ is a surface conformal
to the separating surface $\cal B$ in $\Sigma$.

Let $l,m$ be a pair of normal vectors to $\tilde {\cal B}$
such that $l^2=m^2=1$, $(l\cdot m)=0$. We choose direction of $l$ along a Killing field $\partial_\tau$
which generates
time translations of the bulk manifold $\tilde {\cal M}$. (Let us emphasize
that $\tau$ is a Euclidean time.) Once $l$ is chosen along $\partial_\tau$, $m$ is tangent to $\tilde{\Sigma}$.
Position of $\tilde{\cal B}$ in $\tilde{\Sigma}$ is shown on Fig. \ref{F1}.
We also define 3 unit vectors $n,p,q$ at $\partial \tilde {\cal B}$  which are also tangent to $\tilde{\Sigma}$:
$n$ is orthogonal to $\partial \tilde{\Sigma}$, $q$ is orthogonal to $\partial \tilde {\cal B}$ and tangent to
$\tilde {\cal B}$, $p$ is
orthogonal to $\partial \tilde {\cal B}$ and tangent to $\partial \tilde{\Sigma}$, see Fig. \ref{F1}.

We need  asymptotic relations for the metric on $\tilde{\cal B}$ similar to (\ref{7.1a}). Below
we present a number of results, part of which are new. The details
of computations can be found in Appendices B and C.

Since $\cal M$ is static $\cal B$ has a single
non-trivial extrinsic curvature tensor in $\cal M$ which we denote $k_{\mu\nu}$.
In general, $\tilde{\cal B}$ is tilted to $\partial \tilde{\Sigma}$. That is why
there is a non-vanishing angle between vectors $m$ and $p$. If $\tilde{\cal B}$ is
a minimal surface in asymptotically AdS
space one finds (see Appendix B)  the following asymptotic formula for the tilt angle:
\begin{equation}\label{7.6}
\varphi\simeq {z  \over d-2}~k~~,
\end{equation}
where $k$ is the trace of extrinsic curvature tensor $k_{\mu\nu}$ of $\cal B$.
Surface $\tilde{\cal B}$ becomes orthogonal to $\partial \tilde{\Sigma}$ in the limit $z\to 0$.

The metric induced on $\tilde{\cal B}$ can be written in the form
\begin{equation}\label{7.7}
ds^2(\tilde{\cal B})={1 \over z^2}\left({dz^2 \over \cos^2\varphi}
+ \sigma_{ab}(z,y)dy^a dy^b\right)~~.
\end{equation}
Let $y^a$, $a=1,...d-2$, be coordinates on $\cal B$, and
$\sigma_{ab}(y)=x^\mu_{~,a} g_{\mu\nu}(x) x^\nu_{~,b}$
be the metric induced on $\cal B$ under
embedding $x^\mu=x^\mu(y)$. One finds the following asymptotic formula:
\begin{equation}\label{7.8}
\sigma_{ab}(z,y)=\sigma_{ab}(y)-{z^2 \over d-2}\left(kk_{ab}+R_{ab}-{\sigma_{ab} \over 2(d-1)}R\right)+..,
\end{equation}
where $R_{ab}=x^\mu_{~,a} x^\nu_{~,b} R_{\mu\nu}$,
$k_{ab}=x^\mu_{~,a} x^\nu_{~,b} k_{\mu\nu}$, see Appendix C.

In $d=4$
$$
\mbox{vol}(\tilde{\cal B})=\int_{\cal B}\sqrt{\sigma}d^2y\left[{1 \over 2\varepsilon^2}
+\frac 14 \ln {\mu \over \varepsilon}\left(-R({\cal B})+C_{ijij}+{k^2 \over 2} -
\mbox{Tr}(k^2) \right)+...\right]
$$
\begin{equation}\label{7.9}
={1 \over 2\varepsilon^2}A({\cal B})+{\pi \over 2}(F_a+F_c+F_b)\ln {\mu \over \varepsilon}+...
~~,
\end{equation}
where $\mu$ has a meaning of an infrared cutoff.
This result coincides with computations of \cite{Schwimmer:2008yh}, see also \cite{Hung:2011xb}.

Let us discuss invariant functionals on $\tilde{\cal B}$ which are integrals of curvature invariants.
Since $\tilde{\cal M}$ is a solution to the bulk Einstein equations with the negative cosmological
constant the scalar curvature and the Ricci tensor are known:
$\tilde{R}=-(d+1)d$, $\tilde{R}_{LP}=-d g_{LP}$. These structures are fixed and are
not related to geometrical characteristics of $\cal M$ and $\cal B$.

The other candidate is the Riemann tensor $\tilde{R}_{LPKN}$ of $\tilde{\cal M}$.
In the asymptotic $\tilde{R}_{LPKN}\simeq -(g_{LK}g_{PN}-g_{LN}g_{PK})$ at $z\to 0$. One can also
consider quantities connected with the geometry of $\tilde{\cal B}$. There are two such
quantities: the scalar  curvature
$\tilde{R}_B$ and a quadratic combination
of extrinsic curvature tensors of $\tilde{\cal B}$.
Since $\tilde{\cal B}$ is static the single non-vanishing
extrinsic curvature tensor $K_{LP}$ is defined for the normal vector $m$, see Fig. \ref{F1}.
$K_{LP}$ is traceless since $\tilde{\cal B}$ is minimal.

Thus, one has 3 possible invariants which  behave at $z\to 0$ as follows:
\begin{equation}\label{7.12}
\tilde{R}_{rsrs}(z,y)\equiv  2\tilde{R}_{LPKN}l^L m^P l^K m^N=-2+z^2C_{ijij}(y)+...~~~,
\end{equation}
\begin{equation}\label{7.13}
\tilde{R}_B(z,y)=-(d-2)(d-1)+z^2\left(C_{ijij}(y)+{k^2 \over d-2} -\mbox{Tr}(k^2)\right)+...~~~,
\end{equation}
\begin{equation}\label{7.14}
\mbox{Tr} K^2=K_{LP}K^{LP}=-z^2\left({k^2 \over d-2} -\mbox{Tr}(k^2)\right)+...~~~,
\end{equation}
see calculations in Appendix \ref{A-C}. These asymptotics contain only conformally covariant structures:
normal
projection (\ref{6.13}) of the Weyl tensor of $\cal M$ and combinations of extrinsic curvatures of $\cal B$,
whose transformation properties are listed in Appendix A. The left hand sides of (\ref{7.12})-(\ref{7.14})
are invariant with respect to bulk diffeomorphisms which have a subgroup of so called Penrose-Brown-Henneaux
(PBH) transformations \cite{Brown:1986nw} which preserve the gauge (\ref{7.1a}).
Since PBH transformations generate Weyl
transformations of the metric on $\cal M$, only Weyl covariant structures appear on the right hand sides.

Note that invariants (\ref{7.12})-(\ref{7.14}) are not actually independent:
\begin{equation}\label{GC}
\tilde{R}_B-\tilde{R}+2\tilde{R}_{ss}
-\tilde{R}_{rsrs}-K^2+\mbox{Tr}(K^2)=0
~~.
\end{equation}
Here $\tilde{R}_{ss}=(l^L l^P+m^L m^P)\tilde{R}_{LP}$ and $K=0$ in the considered case.
This identity is the known Gauss-Codazzi equation.

Formulas (\ref{7.9}), (\ref{7.13}) were obtained in \cite{Schwimmer:2008yh} by using
arguments based on properties of PBH transformations.
We derive these and other relations straightforwardly in Appendix \ref{A-C}.

\subsection{Holographic entanglement entropy}

It is instructive first to see how the geometrical structures above work to reproduce the entanglement
entropy in 4D CFT. One uses the holographic formula (\ref{1.1}), AdS/CFT dictionary
which identifies the gravity coupling with the group parameter as $1/G_N^{(5)}=2N^2/\pi$ and finds
the entanglement entropy in form (\ref{1.2}) for the inverse order parameter $\gamma_n=1$. If the gravity
cutoff is chosen as $\varepsilon = \Lambda^{-1}$ one gets relations
\begin{equation}\label{7.10}
s_2^{(1)}={N^2 \over 4\pi}~~,~~
s_4^{(1)}=N^2 \left(\frac 14 F_a+\frac 14 F_c+\frac 14 F_b\right)~~~,
\end{equation}
which coincide exactly with (\ref{6.6}), (\ref{6.8}), (\ref{6.11}), (\ref{6.10}) at large $N$.
If the extrinsic curvature is non-zero one can use (\ref{7.10}) to guess the coefficient
at the invariant $F_b$ which has not been computed on the CFT side so far.
One finds
\begin{equation}\label{7.11}
b(\gamma_n=1)=1/4~~~.
\end{equation}
This `holographic' argument was first suggested in \cite{Solodukhin:2008dh}.

Formula (\ref{7.10}) for $s_4^{(1)}$ does not reproduce a possible contribution of zero modes
$s^{(n=1)}_{\mbox{\tiny{zm}}}$, see (\ref{6.5a}).

\subsection{Holographic R\'{e}nyi entropy}

Our aim now is to find a holographic formula which would be able to reproduce R\'{e}nyi entropy (\ref{1.2})
with coefficients established in Sect. \ref{RCFT}. To be more specific the holographic
R\'{e}nyi entropy $S(n,\tilde{\cal B})$ associated to an entangling surface $\cal B$ in the boundary CFT
is considered to be a functional set on a codimension 2 minimal hypersurface $\tilde{\cal B}$
in $\tilde{\cal M}$. We conjecture that $S(n,\tilde{\cal B})$ is an invariant functional similar to (\ref{1.1})
and is constructed from intrinsic and extrinsic geometrical structures of $\tilde{\cal B}$.

Let us emphasize that the bulk geometries $\tilde{\cal M}$, $\tilde{\cal B}$ are the same as in
the Ryu-Takayanagi setup and
they do not depend, therefore, on the order parameter $n$.
It is the form of the functional which is allowed to contain $n$.
This differs our approach from the recent conjecture of \cite{Hung:2011nu} where $n$ enters the
metric of the bulk solutions.

The simplest case which illustrates our idea is the holographic formula for R\'{e}nyi entropy of a 2D CFT.
One can check that the following area functional:
\begin{equation}\label{7.21a}
S(n,\tilde{\cal B})\simeq {1 \over 4G_N^{(3)}}{\gamma_n+1\over 2} \mbox{vol}(\tilde{\cal B})
\end{equation}
reproduces (\ref{6.6a}) if we choose $c={3 \over 2G_N^{(3)}}$.

To discuss 4D CFT we should consider several invariant structures associated to $\tilde{\cal B}$,
first of all, $\mbox{vol}(\tilde{\cal B})$ and
integrals over $\tilde{\cal B}$ of local invariants constructed from curvatures.
Let us define the following functionals:
\begin{equation}\label{7.15}
\tilde{F}_c={1 \over 2\pi}\int_{\tilde{\cal B}}\sqrt{\tilde{\sigma}}d^{3}y
\left[\tilde{R}_{rsrs}+2l^{-2}\right]~~~,
\end{equation}
\begin{equation}\label{7.16}
\tilde{F}_b=-{1 \over 2\pi}\int_{\tilde{\cal B}}\sqrt{\tilde{\sigma}} d^{3}y~\mbox{Tr} K^2~~~,
\end{equation}
where we denoted  a metric induced on $\tilde{\cal B}$ by $\tilde{\sigma}$ and restored
explicit dependence on the AdS radius $l$.
These quantities are 'holographic duals' of functionals $F_c$ and $F_b$, see (\ref{6.12}), (\ref{6.15}),
in a sense that
\begin{equation}\label{7.17}
\tilde{F}_c=F_c ~l\ln {\mu \over \varepsilon}+...~~~,
\end{equation}
\begin{equation}\label{7.18}
\tilde{F}_b=F_b ~l\ln {\mu \over \varepsilon}+...~~~.
\end{equation}
These results follow directly from (\ref{7.12}) and (\ref{7.14}). Integral of
the scalar curvature $\tilde{R}_B$
according to (\ref{7.13}) is reduced
to a combination of $F_c$ and $F_b$. There is no need to consider this integral independently since
it can be expressed in terms
of (\ref{7.17}) and (\ref{7.18}) with the help of Gauss-Codazzi equality.

To find a holographic representation of the R\'{e}nyi entropy in 4D CFT we also need an invariant functional
'dual' to $F_a$, see (\ref{6.9}). The problem is that $F_a$ is an integral of the scalar
curvature of $\cal B$. This integral is a topological invariant only in $d=4$ when
$\cal B$ has dimension 2. In other dimensions $F_a$ is neither topological nor Weyl invariant.
On the other hand, the PBH transformations require that liner combinations of curvatures in AdS
in any dimension correspond to Weyl invariant structures on the boundary, as in
case of Eqs. (\ref{7.12})-(\ref{7.14}). Therefore,
the bulk functional corresponding to $F_a$ cannot be among local invariants linear in curvatures.
The only local structure which produces $F_a$ is the volume $\mbox{vol}(\tilde{\cal B})$, see
(\ref{7.9}). However, the coefficient by $\mbox{vol}(\tilde{\cal B})$ is fixed by the
leading (area) term in ERE.

By taking this into account one should look for {\it non-local} bulk functionals. For example,
for ERE with entangling surface $\cal B$  having topology of $S^2$ ($F_a=-4$) one can choose
\begin{equation}\label{7.23}
\tilde{F}_a=-2l \ln (\mbox{vol}(\tilde{\cal B})/ l^3)~~.
\end{equation}
Equation (\ref{7.9}) can be used to show that
\begin{equation}\label{7.24}
\tilde{F}_a=F_a~l\ln {\mu \over \varepsilon}-2l\ln (A({\cal B})/ \mu^2)+...~~~,
\end{equation}
Certainly, (\ref{7.23}) is not a single option and other non-local structures are possible.
The choice of (\ref{7.23}) seems to be the simplest. The logarithmic term in a holographic ERE
is similar to logarithmic corrections to the Bekenstein-Hawking entropy, which
are a rather common consequence of quantum effects, see e.g. \cite{Fursaev:1994te}-\cite{Sen:2011ba}.

Now one can check with the help of (\ref{7.9}), (\ref{7.17}), (\ref{7.18}), and (\ref{7.24})
that the following functional:
\begin{equation}\label{7.21}
S(n,\tilde{\cal B})={1 \over 4G_N^{(5)}}\left(\gamma_n\mbox{vol}(\tilde{\cal B})+
2\pi l^2\left(\tilde{a}(\gamma_n)\tilde{F}_a+
\tilde{c}(\gamma_n)\tilde{F}_c+\tilde{b}(\gamma_n)\tilde{F}_b\right)\right)~~,
\end{equation}
\begin{equation}\label{7.22a}
\tilde{a}(\gamma_n)=a(\gamma_n)-\frac 14\gamma_n={1 \over 96}(\gamma_n-1)\left(\gamma_n^2+2\gamma_n-15\right)~~,
\end{equation}
\begin{equation}\label{7.22c}
\tilde{c}(\gamma_n)=c(\gamma_n)-\frac 14\gamma_n={1 \over 32}(\gamma_n-1)\left(\gamma_n^2+2\gamma_n-3\right)~~,
\end{equation}
\begin{equation}\label{7.22b}
\tilde{b}(\gamma_n)=b(\gamma_n)-\frac 14\gamma_n~~,
\end{equation}
reproduces (\ref{6.6}), (\ref{6.8}), (\ref{6.11}), (\ref{6.10}) at large $N$.
Adding $-\frac 14\gamma_n$ in  (\ref{7.22a})-(\ref{7.22b}) compensates
curvature terms ($F_a$,$F_b$, $F_c$) which come out from $\mbox{vol}(\tilde{\cal B})$, see
(\ref{7.9}).

Since $\tilde{a}(1)=\tilde{c}(1)=\tilde{b}(1)=0$ expression (\ref{7.21})
coincides with Ryu-Takayanagi formula (\ref{1.1}) in the limit $n\to 1$.
It is also important to note that $S^{(n)}({\cal B})$, like (\ref{1.1}), depends only on
low-energy constants $G_N^{(5)}$ and $l$ which enter gravity action (\ref{7.1}).

\section{Discussion}\label{spec}
\setcounter{equation}0

We suggest (\ref{7.21}) as a holographic formula for the entanglement R\'{e}nyi entropy.
One should emphasize that $S(n,\tilde{\cal B})$ reproduces
at least the divergent part (\ref{1.2}) of ERE
in the conformal theory in four dimensions. Thus, additional terms may be needed in $S(n,\tilde{\cal B})$
to go beyond the given approximation.

One may note  some differences between (\ref{7.21})
and Ryu-Takayanagi expression (\ref{1.1}).
Unlike (\ref{1.1}), the form of $S(n,\tilde{\cal B})$
explicitly depends on the dimension $d$. This can be seen
by comparing (\ref{7.21}) with
a possible formula of holographic ERE for 2D CFT, see (\ref{7.21a}).
In (\ref{7.21a}) the leading term
remains finite in the limit $n\to \infty$ while in (\ref{7.21}) the volume term vanishes.

Another distinction between (\ref{1.1}) and (\ref{7.21}) is that
the minimal hypersurface $\tilde{\cal B}$ is not an exact extremal
hypersurface for functional (\ref{7.21}). (An extremal value of (\ref{7.21}) is defined
by varying position of the hypersurface under fixed background metric and $n$.)
Note that $\tilde{\cal B}$ would be an
extremal hypersurface if $\tilde{F}_a$ term alone were present.
This term depends only on the volume $\mbox{vol}(\tilde{\cal B})$. However
$S(n,\tilde{\cal B})$ includes also terms with $\tilde{F}_b$ and  $\tilde{F}_c$ which
depend on curvatures.  Let us emphasize that the
choice of $\tilde{\cal B}$ as an argument in the holographic ERE  functional
was crucial for finding the correspondence between
bulk and boundary quantities.

What happens if $\tilde{\cal B}$ is replaced by a genuine extremal surface?
Let us define (\ref{7.21}) on a set of codimension 2 hypersurfaces $\tilde{\cal Q}$ in $\tilde{\cal M}$
specified by the same boundary condition as $\tilde{\cal B}$, that is
$\partial\tilde{\cal Q}\sim {\cal B}$.
Such hypersurfaces have infinitely large volume in the limit $z \to 0$.
We also require that for $\tilde{\cal Q}$ the following
restrictions are satisfied at $z\to 0$: $\mbox{Tr}~ K^2 \ll l^{-2}$ and $|\tilde{R}_{rsrs}+2l^{-2}|\ll l^{-2}$.
These restrictions allow $\tilde{\cal B}$ to belong to the given set.
They also guarantee that terms with $\tilde{F}_b$ and  $\tilde{F}_c$ in functional $S(n,\tilde{\cal Q})$
are small compared to the volume term. To see this one should take into account that
main contributions to integrals are picked up in a neighborhood of $z=0$.

Suppose that $S(n,\tilde{\cal Q})$ has an extremum on some hypersurface $\tilde{\cal Q}$
from the considered set. Extremal hypersurfaces may depend on $n$, in general.
Since terms with $\tilde{F}_b$ and  $\tilde{F}_c$ are small one can represent $\tilde{\cal Q}$
as $\tilde{\cal B}$ with a small perturbation, $\tilde{\cal Q}=\tilde{\cal B}+\tilde{\cal Q}_1$.
Up to terms which are of the second order in the perturbation,
$S(n,\tilde{\cal Q})\simeq S(n,\tilde{\cal B})$.
The second order terms are given by some (non-local) functional on $\tilde{\cal B}$ quadratic in curvatures
which enter $\tilde{F}_b$ and  $\tilde{F}_c$.
It is important that the second order terms are small compared to $S(n,\tilde{\cal B})$. Therefore,
once $S(n,\tilde{\cal B})$ is used to reproduce ERE only in the logarithmic
approximation it is safe to replace it with the extremal value $S(n,\tilde{\cal Q})$.

The entanglement R\'{e}nyi entropy $S^{(n)}({\cal B})$ can be defined by formula (\ref{2.4})
in terms of an entanglement partition function $Z(n)$. It is interesting to discuss if there is a
holographic representation for  $Z(n)$  which results in functional $S(n,\tilde{\cal B})$.
We use a line of reasonings suggested in \cite{Fursaev:2006ih} to show that such a representation
may be possible.

In a quantum field theory  $Z(n)$ can be written in terms of a path integral
where field configurations are located on a manifold ${\cal M}_n$ glued from $n$ copies of the physical
manifold $\cal M$
along some cuts $\Sigma_A$ which meet on $\cal B$, see \cite{Fursaev:2006ng}.
($\cal B$ divides a constant time hypersurface in $\cal M$ on parts $\Sigma_A$ and $\Sigma_B$, the
trace in the reduced density matrix (\ref{2.6}) is taken over states
on $\Sigma_B$.) The AdS/CFT correspondence implies that $Z(n)$ can be replaced by a partition function,
$Z^{AdS}[{\cal M}_n]$, in AdS-gravity for which a given CFT is a 'boundary' theory.
The idea of \cite{Fursaev:2006ih} is that in a low-energy approximation one should look
for a path integral representation of $Z^{AdS}[{\cal M}_n]$
with the condition that the conformal boundary of "histories", $\tilde{\cal M}_n$, involved
in the path integral belongs to the conformal class of ${\cal M}_n$,
\begin{equation}\label{8.2}
Z^{AdS}[{\cal M}_n]=\int_{\tilde{\cal M}_n:~\partial \tilde{\cal M}_n \sim {\cal M}_n} [Dg]\exp(-W[g])~~~.
\end{equation}
Functional $W[g]$ is an effective action which is induced by quantum gravity or string theory
dynamics in the AdS bulk.
For regular boundary conditions $W[g]$ is approximated by classical action (\ref{7.1}).
Since the boundary manifolds have conical singularities application of (\ref{7.1}) in this case
is not obvious.

Although quantum gravity arguments are absent we make a suggestion for $W[g]$
consistent with the holographic ERE. For the given boundary conditions
there are two types of geometries $\tilde{\cal M}_n$ which may contribute to (\ref{8.2})
in a semiclassical approximation.
One type includes spaces which are regular in the bulk, another
type includes manifolds with conical singularities.  We consider only singular geometries.
They can be constructed for the given boundary conditions
in the following way \cite{Fursaev:2006ih}.
One starts from bulk geometries $\tilde{\cal M}$
such that $\partial \tilde{\cal M}\sim {\cal M}$.
Then, one makes different cuts of $\tilde{\cal M}$ along
$d$--dimensional hypersurfaces $\tilde{\Sigma}_A$ with the boundary condition
$\partial \tilde{\Sigma}_A \sim \Sigma_A$, where $\Sigma_A$ is the corresponding cut in ${\cal M}$.
The boundary condition does not fix $\tilde{\Sigma}_A$ uniquely.
By taking $n$ identical copies of $\tilde{\cal M}$ with the same cut
$\tilde{\Sigma}_A$ and gluing them along the cuts one gets a space $\tilde{\cal M}_n$ with the
required boundary condition $\partial \tilde{\cal M}_n \sim {\cal M}_n$.
The bulk spaces have conical singularities on codimension
2 hypersurfaces $\tilde{\cal Q}$ related to the entangling surface,
$\partial \tilde{\cal Q} \sim \cal B$.
The holographic ERE defined by AdS partition function and approximated by using (\ref{8.2}) is
\begin{equation}\label{8.8}
S^{AdS}(n)= {1 \over 1-n}\left(\ln Z^{AdS}[{\cal M}_n]-n\ln Z^{AdS}[{\cal M}]\right)
\simeq{1 \over n-1}\left(W(n)-nW(1)\right)
\end{equation}
where $W(n)$ is a least value of the effective action on some singular space $\tilde{\cal M}_n$.

We want to find $W(n)$ by requiring that $S^{AdS}(n)\simeq S(n,{\cal B})$.
It is natural to assume that the leading part of $W(n)$ is local and
is similar to a divergent part of a QFT effective action on manifolds with conical
singularities. The structure of the heat kernel coefficients (\ref{6.17}) then implies that
$W(n)$ is a sum of two terms: one is defined on a regular domain $\tilde{\cal M}_n/\tilde{\cal Q}$,
the other is located on $\tilde{\cal Q}$ and induced by quantum effects on conical singularities.
The term on a regular domain coincides with classical
action (\ref{7.1}). It is proportional to $n$ and does not contribute to ERE (\ref{8.8}).
The only possible form of the leading part of $W(n)$ which allows one to equate the two entropies,
$S^{AdS}(n)\simeq S(n,\tilde{\cal Q})$, is
\begin{equation}\label{8.7}
W(n)\simeq I[\tilde{\cal M}_n/\tilde{\cal Q}]+(n-1)
S(n,\tilde{\cal Q})~~.
\end{equation}
What is a singular manifold where the effective action has a least value?
The fact that a saddle 'point' is a singular manifold even without a matter source
which supports the conical singularities should not be considered as a controversy.
The extremum
of $W(n)$ is defined within a restricted set of geometries. For other class of geometries
which contribute to (\ref{8.2}) and are regular in the bulk one should use another action.
One should also note that
(\ref{8.7}) does not coincide with a classical action (\ref{7.1}) naively taken on $\tilde{\cal M}_n$.
In this case one would not get in (\ref{8.7}) $\tilde{F}_a$, $\tilde{F}_b$ and $\tilde{F}_c$ terms.
The two actions agree only in the limit $n\to 1$ up to terms linear in $(n-1)$.

Variations of the two terms in (\ref{8.7}) are required to vanish independently.
Variations of $I[\tilde{\cal M}_n/\tilde{\cal Q}]$ are subject to certain boundary conditions near
conical singularities to preserve their structure. They result
in the standard bulk gravity equations for (\ref{7.1}). That is, locally $\tilde{\cal M}_n$ is one
of solutions of AdS gravity. Minimization of $S(n,\tilde{\cal Q})$ implies that $\tilde{\cal Q}$
is an extremal hypersurface, thus, in the leading approximation we recover
the holographic entanglement entropy $S(n,\tilde{\cal B})$ from $S^{AdS}(n)$.

Although the above 'derivation' of holographic ERE is based on a number of assumptions it
may be a plausible scenario. An explanation of Ryu-Takayanagi formula (\ref{1.1}) seems to
be its particular case which follows in the limit $n\to 1$. This gives a further support to earlier
arguments presented in \cite{Fursaev:2006ih} and allows one to avoid their criticism in
\cite{Headrick:2010zt}.

It would be very interesting to understand the behaviour of ERE in a strong coupling 
regime. Our approach to the holographic description of ERE should hold in this 
case but functions $\tilde{a}(\gamma_n)$, 
 $\tilde{b}(\gamma_n)$,  $\tilde{c}(\gamma_n)$ in (\ref{7.21}) may be different.
The reason is that the logarithmic terms in the Renyi entropy for $n>1$ are not determined solely
by the conformal anomaly. They are not protected from both perturbative and non-perturbative
corrections, as was pointed out in \cite{Klebanov:2011uf}. 
This may be the reason why results of \cite{Hung:2011nu} for a
spherical entangling surface disagree with weak coupling formula
(\ref{rel3}). 

Our proposal may be compatible with the approach of \cite{Hung:2011nu} (after a proper
redefinition of  $\tilde{a}(\gamma_n)$, $\tilde{b}(\gamma_n)$, and  $\tilde{c}(\gamma_n)$) but 
to resolve the issue one needs to know quantum corrections to ERE.

\newpage
\appendix
\section{Conformal transformations}\label{A-A}
\setcounter{equation}0

Several useful relations for conformal transformations of the metric of a $D$ dimensional manifold
\begin{equation}\label{a1.1}
\bar{g}_{\mu\nu}(x)=e^{-2\omega(x)}g_{\mu\nu}(x)~~
\end{equation}
are listed below for the sake of completeness.
Dimensionality $D$ in cases considered in our work is either $d+1$ or $d$, where $d$ is the dimensionality
of the boundary CFT. One can find the following transformations:
$$
\bar{R}_{\lambda\mu\nu\rho}=e^{-2\omega}\left[R_{\lambda\mu\nu\rho}+\omega_{\lambda\nu}g_{\mu\rho}
-\omega_{\lambda\rho}g_{\mu\nu}-\omega_{\mu\nu}g_{\lambda\rho}-\omega_{\mu\rho}g_{\lambda\nu}+\right.
$$
\begin{equation}\label{a1.2}
\left. \omega_{\mu}\omega_{\rho}g_{\lambda\nu}-\omega_{\nu}\omega_{\mu}g_{\lambda\rho}+
\omega_{\nu}\omega_{\lambda}g_{\rho\mu}-\omega_{\lambda}\omega_{\rho}g_{\mu\nu} +
\omega^\alpha\omega_\alpha(g_{\mu\nu}g_{\lambda\rho}-g_{\mu\rho}g_{\lambda\nu})\right]~~,
\end{equation}
\begin{equation}\label{a1.3}
\bar{R}_{\mu\nu}=R_{\mu\nu}+(D-2)\omega_{\mu\nu}+g_{\mu\nu}\Delta\omega+(D-2)\omega_{\mu}\omega_{\nu}+
(2-D)\omega^\alpha\omega_\alpha~g_{\mu\nu}~~,
\end{equation}
\begin{equation}\label{a1.4}
\bar{R}=e^{2\omega}\left[R+2(D-1)\Delta\omega-(D-2)(D-1)\omega^\alpha\omega_\alpha\right]~~,
\end{equation}
where $\omega_\mu=\omega_{,\mu}$, $\omega_{\mu\nu}=\omega_{;\mu\nu}$.
If there is a codimension 2 hypersurface with two normal
vectors $n_i$ one can also establish the following scaling properties:
\begin{equation}\label{a1.1n}
\bar{n}_i^\mu=e^{\omega}n_i^\mu~~,
\end{equation}
\begin{equation}\label{a1.5}
(\bar{k_i})_{\mu\nu}=e^{-\omega}\left[(k_i)_{\mu\nu}-h_{\mu\nu} n_i^\lambda\omega_{,\lambda}\right]~~,
\end{equation}
\begin{equation}\label{a1.6}
\bar{C}_{ijij}=e^{2\omega}C_{ijij}~~,
\end{equation}
\begin{equation}\label{a1.7}
\bar{k}_i^2-(D-2)\mbox{Tr}(\bar{k}_i^2)=e^{2\omega}\left(k_i^2-(D-2)\mbox{Tr}(k_i^2)\right)
~~,
\end{equation}
where $h_{\mu\nu}=g_{\mu\nu}-(n_i)_\mu (n_i)_\nu$ is the metric on the hypersurface,
$(k_i)_{\mu\nu}=h_{\mu}^{\lambda}h_{\nu}^{\rho}(n_i)_{\mu;\nu}$ are
its extrinsic curvatures, $C_{ijij}$ is the normal projection (\ref{6.13}) of the Weyl tensor
(\ref{6.14}). Our convention is that $n_i$ are outward pointing vectors.

\section{Tilt angle of the holographic surface in AdS}\label{A-B}
\setcounter{equation}0

To prove (\ref{7.6}) for the tilt angle $\varphi$ shown on Fig. \ref{F1} we consider embedding of
$\tilde{\cal B}$ described by the equation
\begin{equation}\label{A-B.1}
z=\psi(x)~~~.
\end{equation}
From definition of vectors $m$ and $n$ one finds
\begin{equation}\label{A-B.2}
\sin\varphi=(m\cdot n)={1 \over \sqrt{1+(\psi')^2}}~~~,
\end{equation}
where $(\psi')^2=\psi_{,\mu}g^{\mu\nu}\psi_{,\nu}$ and $g_{\mu\nu}$ 
is defined in (\ref{7.1a}). We suppose that $\tilde{\cal B}$ is a minimal
codimension 2 hypersurface
embedded in $d+1$ dimensional manifold which is a solution to the Einstein equations
with a negative cosmological constant. In this case $\psi$ is a solution to the equation
\begin{equation}\label{A-B.3}
{d-1 \over \psi\sqrt{1+(\psi')^2}}+{1 \over \sqrt{g}}\partial_\mu
\left(\sqrt{g}{g^{\mu\nu}\psi_{,\nu} \over \sqrt{1+(\psi')^2}}\right)=0~~~
\end{equation}
which determines a minimum of $\mbox{vol}(\tilde{\cal B})$.

We prove (\ref{7.6}) by studying asymptotic of the tilt angle in the form $\varphi\simeq \lambda z$.
The parameter $\lambda$ is fixed for known examples, when $\partial \tilde{\cal B}$
(or $\cal B$) is either a hypersphere, 2 infinite parallel planes or a cylinder.
The bulk space $\tilde{\cal M}$ is assumed to be a pure AdS space.

1. {\it Sphere}. We choose the $d$-dimensional part of metric (\ref{7.1a}) in the form
\begin{equation}\label{A-B.4}
g_{\mu\nu}dx^\mu dx^\nu=d\tau^2+dr^2+r^2d\Omega^2_{d-2}~~~,
\end{equation}
$d\geq 3$, where $d\Omega^2_{d-2}$ is a metric on a unit sphere $S^{d-2}$. The surface ${\cal B}$ is
$S^{d-2}$ with the radius $r=R$. The trace of extrinsic curvature of ${\cal B}$ is $k=(d-2)/R$.
Equation for $\tilde{\cal B}$ is $\psi=\psi(r)$.
The solution to (\ref{A-B.3}) for the given boundary condition is \cite{Ryu:2006ef}
\begin{equation}\label{A-B.5}
\psi(r)=\sqrt{C^2-r^2}~~~,
\end{equation}
where $C=\sqrt{z^2+R^2}$ and $z=const$ is a position of $\partial \tilde{\cal B}$.
One has from (\ref{A-B.2}) and (\ref{A-B.5})
\begin{equation}\label{A-B.6}
\sin\varphi={\psi \over C}\simeq {z \over R}~~~,
\end{equation}
where we took into account that $C=R$ in the limit $z\to 0$. Eq. (\ref{A-B.6}) is equivalent
to (\ref{7.6}) at small $z$.

2. {\it Parallel planes}. The metric is
\begin{equation}\label{A-B.7}
g_{\mu\nu}dx^\mu dx^\nu=d\tau^2+dy^2+dx_1^2+...+dx_{d-2}^2~~~,
\end{equation}
$d\geq 3$. The surface ${\cal B}$ consists of 2 parallel planes with positions $y=0$, $y=a$.
The extrinsic curvatures of ${\cal B}$ are equal to zero. Equation for $\tilde{\cal B}$ is $\psi=\psi(y)$.
The solution to (\ref{A-B.3}) has the property
\cite{Ryu:2006ef}
\begin{equation}\label{A-B.8}
\psi'_y(y)=\psi^{-(d-1)}\sqrt{C^2-\psi^{2(d-1)}}~~~,
\end{equation}
where $C$ is a constant which is fixed by the boundary conditions.
One finds from (\ref{A-B.2}) and (\ref{A-B.8}) that
\begin{equation}\label{A-B.9}
\sin\varphi={\psi^{(d-1)} \over C}\sim z^{(d-1)}~~~,
\end{equation}
where we took into account that $C=R$ in the limit $z\to 0$.
Thus, $\varphi=O(z^{d-1})$ rather than $O(z)$, as in (\ref{7.6}),
in accord with the fact that $\cal B$ has zero extrinsic curvature.

3. {\it Cylinder}. The metric is
\begin{equation}\label{A-B.10}
g_{\mu\nu}dx^\mu dx^\nu=d\tau^2+d\rho^2+\rho^2 d\theta^2+...+dx_{d-3}^2~~~,
\end{equation}
$d\geq 4$, $0\leq \theta <2\pi$. Equation of $\cal B$ is $\rho=a$, the trace of extrinsic curvature of
${\cal B}$ is $k=1/a$. Equation for $\tilde{\cal B}$ is $\psi=\psi(\rho)$.
The solution to (\ref{A-B.3}) has the following asymptotic near
the boundary \cite{Solodukhin:2008dh}:
\begin{equation}\label{A-B.11}
\psi(\rho)\simeq \sqrt{2a(d-2)(a-\rho)}~~~,
\end{equation}\
which yields for  (\ref{A-B.2})
\begin{equation}\label{A-B.12}
\sin\varphi\simeq{\psi \over a(d-2)}={z \over a(d-2)}~~~.
\end{equation}
Eq. (\ref{A-B.12}) is equivalent
to (\ref{7.6}) for the given extrinsic curvature.

\section{Asymptotic equations of the holographic surface}\label{A-D}
\setcounter{equation}0

In this section we derive Eqs. (\ref{7.7}), (\ref{7.8}). We describe embedding of
$\tilde{\cal B}$ by the equations
\begin{equation}\label{A-D.1a}
x^\mu=x^\mu(z,y)~~,~~\mu\neq \tau,
\end{equation}
\begin{equation}\label{A-D.1b}
~~\tau=const~~,
\end{equation}
where $\tau$ is a time coordinate. Coordinates on  $\tilde{\cal B}$
are $z$ and $y^a$, $a=1,...,d-2$. At constant and small $z$ functions $x^\mu(z,y)$ describe
embedding $x^\mu=x^\mu(y)$ of the boundary  $\partial \tilde{\cal B}$, and, consequently,
embedding of $\cal B$. In this case $y^a$ are coordinates on $\partial \tilde{\cal B}$ and $\cal B$.
One can consider the following decomposition near the boundary:
\begin{equation}\label{A-D.7}
x^\mu(z,y)=x^\mu(y)+x^\mu_{(1)}(y)z+x^\mu_{(2)}(y)z^2+...,~~\mu\neq \tau~~.
\end{equation}
It is convenient to define the vector $q$ in the tangent space to $\tilde{\cal B}$, see Fig. \ref{F1},
\begin{equation}\label{A-D.3}
q=n\cos\varphi-p\sin\varphi ~~,
\end{equation}
where $\varphi$ is the tilt angle. Vector $q$ is unit and orthogonal
(in the tangent space) to the boundary  $\partial \tilde{\cal B}$. To determine subleading
terms in decomposition (\ref{A-D.7}) we note that vectors $\partial_z x^L(z,y)$,
$\partial_a x^L(z,y)$ (where $x^L=(z,x^\mu)$) are in the tangent space and obey the following properties:
$\partial_z x^L(z,y)$ is directed along $q$, while $\partial_a x^L(z,y)$
is orthogonal to $q$. These properties result in conditions:
\begin{equation}\label{A-D.8}
\partial_z x^\mu(z,y)=-\tan \varphi~ \bar{p}^\mu~~,
\end{equation}
\begin{equation}\label{A-D.9}
\bar{p}^\mu g_{\mu\nu}(z,y) \partial_ax^\nu(z,y)=0~~,
\end{equation}
where $g_{\mu\nu}(z,y)/z^2$ is a metric induced on $\partial \tilde{\cal M}$.
In (\ref{A-D.8}), (\ref{A-D.9})  we used  the fact that $n_\mu=p_z=0$ and introduced the vector
$\bar{p}_\mu=zp_\mu$, $\bar{p}^\mu =g^{\mu\nu}\bar{p}_\nu$, which is normalized as
$\bar{p}^\mu\bar{p}_\mu=1$. In the limit $z\to 0$ the metric  $g_{\mu\nu}$ coincides with the
metric on $\cal M$, and $\bar{p}_\mu$ is the normal vector to $\cal B$ in $\cal M$.
It follows from (\ref{7.6}) and integration of (\ref{A-D.8}) over $z$ that
\begin{equation}\label{A-D.10}
x^\mu(z,y)=x^\mu(y)-{k \over 2(d-2)}\bar{p}^\mu ~z^2+...,~~\mu\neq \tau~~.
\end{equation}
Eq. (\ref{A-D.10}) was derived in \cite{Schwimmer:2008yh} by using PBH transformations.

Let us now derive the metric induced on $\tilde{\cal B}$, see Eqs. (\ref{7.7}), (\ref{7.8}).
The metric can be written as
\begin{equation}\label{A-D.2}
ds^2(\tilde{\cal B})=h_{zz}dz^2+2h_{za}dzdy^a+h_{ab}dy^ady^b~,
\end{equation}
\begin{equation}\label{A-D.4a}
h_{zz}={1 \over z^2}(1+\partial_z x^\mu g_{\mu\nu}\partial_z x^\nu)=
{1 \over z^2}(1+\tan^2 \varphi~ \bar{p}^\mu\bar{p}_\mu)=
{1 \over z^2\cos^2 \varphi}~~,
\end{equation}
\begin{equation}\label{A-D.4b}
h_{za}={1 \over z^2}\partial_z x^\mu g_{\mu\nu}\partial_a x^\nu=0~~,
\end{equation}
\begin{equation}\label{A-D.4c}
h_{ab}={1 \over z^2}g_{\mu\nu}\partial_a x^\mu\partial_b x^\nu\equiv {\sigma_{ab} \over z^2}~~,
\end{equation}
where we used (\ref{A-D.8}), (\ref{A-D.9}). Thus, (\ref{A-D.2}) reproduces (\ref{7.7}).
To proceed with computation of $\sigma_{ab}$
near the boundary and prove (\ref{7.8}) one should use (\ref{A-D.9}), (\ref{A-D.10})
\begin{equation}\label{A-D.10a}
\partial_a x^\mu(z,y)=\partial_a x^\mu(y)-{z^2 \over 2(d-2)}\partial_a(k \bar{p}^\mu)+...,
\end{equation}
take into account the definition of the extrinsic curvature of $\cal B$,
\begin{equation}\label{A-D.11}
k_{ab}=x^\mu_{~,a} x^\nu_{~,b}\bar{p}_{\mu;\nu}~~,
\end{equation}
and Fefferman-Graham asymptotic  (\ref{7.2}), (\ref{7.3}).

\section{Asymptotic properties of curvature invariants}\label{A-C}
\setcounter{equation}0

The aim of this section is to prove Eqs. (\ref{7.12}) and (\ref{7.13}).
Equation (\ref{7.14}) follows from (\ref{7.12}), (\ref{7.13}) and Gauss-Codazzi identity
(\ref{GC}). We use results of
Sec. \ref{A-A} and make a conformal transformation
of metric (\ref{7.1a}) to metric
\begin{equation}\label{A-C.1}
ds^2=dz^2 + g_{\mu\nu}(z,x)dx^\mu dx^\nu\equiv g_{KL}dx^K dx^L~~.
\end{equation}
We denote a manifold with metric (\ref{A-C.1}) as $\tilde{\cal M}'$. The holographic
surface after the conformal transformation is mapped to a codimension 2 hypersurface $\tilde{\cal B}'$
in $\tilde{\cal M}'$. Note that  $\partial \tilde{\cal B}'={\cal B}$. The scalar curvature of
$\tilde{\cal B}'$ is $\tilde{R}'_B$.

After the conformal transformation we use the same letters for the normal vectors
shown on Fig. \ref{F1}. The transformation does not change angle $\varphi$.
As earlier $\tilde{\cal B}$ is assumed to be minimal so we can use results of Sec. \ref{A-B}.
We enumerate normal vectors
to $\tilde{\cal B}'$ (vectors $m$ and $l=\partial_\tau$ normalized to unity with the help of (\ref{A-C.1}))
by letters $r,s$. Then, for example, the normal projection of the Riemann
tensor of $\tilde{\cal M}'$ at $\tilde{\cal B}'$ is $\tilde{R}'_{rsrs}=2\tilde{R}'_{KLMN}m^Kl^Lm^Ml^N$.

The letters $i,j$ correspond to normal vectors to ${\cal B}$ (these are vectors $p$ and $l$). Analogously,
repeated indexes $i$ or $j$ denote projection with the help of $p$ and $l$ of tensor components at ${\cal B}$.

One has the relations which follow from (\ref{a1.2}), (\ref{a1.4})
\begin{equation}\label{A-C.2}
\tilde{R}_{rsrs}
=z^2\left(\tilde{R}'_{rsrs}+
2(\omega_{ss}-\omega_{\|}\omega^{\|})\right)~~~,
\end{equation}
\begin{equation}\label{A-C.3}
\tilde{R}_B=z^2\left(\tilde{R}'_B+2(d-2)\Delta_{\|}\omega-(d-2)(d-1)\omega_{\|}\omega^{\|}\right)~~~.
\end{equation}
Here $\omega=\ln z$, while $\omega_{\|}$, $\Delta_{\|}\omega$ are, respectively,
derivatives and the Laplace operator in the space tangent to $\tilde{\cal B}'$.

One finds, by the definition,
$$
\omega_{ss}=(l^Kl^N+m^Km^N)\omega_{;KN}=
$$
\begin{equation}\label{A-C.4}
(l^Kl^N+\sin^2\varphi~ n^Kn^N+2\sin\varphi\cos\varphi~ n^Kp^N
+\cos^2\varphi~ p^Kp^N)\omega_{;KN}\simeq g^{(1)}_{ii}-{k^2 \over (d-2)^2}~~,
\end{equation}
where $n=\partial_z$. The last equality follows in the limit $z\to 0$ if one uses (\ref{7.6})
and the Fefferman-Graham asymptotic (\ref{7.2}). Note that metric $g_{KL}$ is static. In (\ref{A-C.4})
\begin{equation}\label{A-C.5}
g^{(1)}_{ii}=(l^\mu l^\nu+p^\mu p^\nu)g^{(1)}_{\mu\nu}=-{1 \over d-2}\left(R_{ii}-{R \over d-1}\right)~~~,
\end{equation}
see (\ref{7.3}), where $R$ and $R_{\mu\nu}$ are the curvatures of $\cal M$.
In a similar way one finds
\begin{equation}\label{A-C.6}
\omega_{\|}\omega^{\|}=q^Kq^L~\omega_{,K}\omega_{,L}=\cos^2\varphi~ n^Kn^L\omega_{,K}\omega_{,L}
\simeq {1 \over z^2}-{k^2 \over (d-2)^2}~~~.
\end{equation}
Equation (\ref{7.12}) follows from (\ref{A-C.2}), (\ref{A-C.4})-(\ref{A-C.6}) if one takes into account
that in the limit $z\to 0$
\begin{equation}\label{A-C.7}
\tilde{R}'_{rsrs}\simeq 2\tilde{R}'_{KLMN}p^Kl^Lp^Ml^N=R_{ijij}~~~,
\end{equation}
and definition (\ref{6.14}) of the Weyl tensor in the corresponding dimensionality.

To get (\ref{7.13}) one should use (\ref{7.7}), (\ref{7.8}). If one uses notations
\begin{equation}\label{A-C.9}
\sigma_{ab}(z,y)=\sigma_{ab}(y)+z^2 \sigma^{(1)}_{ab}(y)+..,
\end{equation}
\begin{equation}\label{A-C.10}
\sigma^{(1)}_{ab}(y)=-{1 \over d-2}\left(kk_{ab}+R_{ab}-{\gamma_{ab} \over 2(d-1)}R\right)~~~,
\end{equation}
then
\begin{equation}\label{A-C.11}
\tilde{R}'_B(z,y)=R({\cal B})-2\sigma^{(1)}_{ab}\sigma^{ab}+...
\end{equation}
\begin{equation}\label{A-C.12}
\Delta_{\|}\omega=-{1 \over z^2}+\sigma^{(1)}_{ab}\sigma^{ab}+...~~~.
\end{equation}
Equation (\ref{7.13}) follows from (\ref{A-C.3}),(\ref{A-C.6}),(\ref{A-C.11}), and (\ref{A-C.12}).
One should also take into account an analog of Gauss-Codazzi relation (\ref{GC})
for $\cal B$ in $\cal M$ and the definition of the Weyl tensor.

\end{document}